\documentclass[10pt]{iopart}

\usepackage[english]{babel}
\usepackage{iopams}
\usepackage[dvips]{color}
\usepackage{graphicx}

\begin{document}

\title[Transformation kinetics of alloys]{Transformation
kinetics of alloys under non-isothermal conditions}

\author{A. R. Massih$^{1,2}$\footnote{Corresponding author.} and L. O. Jernkvist$^1$}

\address{
$^1$Quantum Technologies, Uppsala Science Park, SE-751 83 Uppsala and \\
$^2$Malm\"{o} University, SE-205 06 Malm\"{o}, Sweden}
\ead{alma@quantumtech.se}

\begin{abstract}
  The overall solid-to-solid phase transformation kinetics under
non-isothermal conditions has been modelled by means of a differential
equation method.  The method requires provisions for expressions of
the fraction of the transformed phase in equilibrium condition and the
relaxation time for transition as functions of temperature.  The
thermal history is an input to the model. We have used the method to
calculate the time/temperature variation of the volume fraction of the
favoured phase in the $\alpha\Leftrightarrow\beta$ transition in a
zirconium alloy under heating and cooling, in agreement with
experimental results. We also present a formulation that accounts for both
additive and non-additive phase transformation processes. Moreover, a method
based on the concept of path integral, which considers all
the possible paths in thermal histories to reach the final state, is
suggested.
\end{abstract}


\section{Introduction}
\label{sec:intro}
The kinetics of phase transformations in solids often involves the
effects of heating and/or cooling rates. This is because first order
phase transformations generally occur by concurrent nucleation and
growth of the new phase and that both these mechanisms are time and
temperature dependent \cite{Gunton_et_al_1983,Langer_1992}. The
overall phase transformation kinetics (nucleation plus growth) is
represented by the fraction of transformed material in the system as a
function of time and temperature $\phi=\phi(t,T)$.  This attribute
under isothermal conditions will results in a
time-temperature-transformation (TTT) diagram, which is a complement
to the phase diagram in equilibrium thermodynamics. Besides the
nucleation rate and the growth rate, other factors that determine
$\phi$ include the density and distribution of nucleation sites, the
overlap of diffusion fields from neighbouring transformed volumes, and
the collision by adjacent transformed domains
\cite{Porter_Easterling_1981}.

 A phenomenological stochastic kinetic model for the overall phase
transformation under \emph{isothermal} conditions was solved exactly by
Kolmogorov \cite{Kolmogorov_1937} and later independently by
Johnson and Mehl \cite{Johnson_Mehl_1939} and Avrami \cite{Avrami_1939}, the KJMA
model.  Early treatments of non-isothermal transformation kinetics are
found in papers by Avrami \cite{Avrami_1940} and
Cahn \cite{Cahn_1956b}. Avrami \cite{Avrami_1940} showed that for a
special case where the nucleation rate is proportional to the the
growth rate of the favoured phase, over a temperature range, the
non-isothermal transformations can be considered as a series of
isothermal reactions at every time step that can be linearly
superposed to give the \emph{non-isothermal} condition (additivity
rule). Later, Cahn \cite{Cahn_1956b} argued that phase transformations
that involve heterogeneous nucleation quite often obey an additivity
rule. He noted that for such situations, a non-isothermal
transformation can be related to an isothermal one through simple rate
rules. More specifically, Cahn showed that if the transformation rate
(${\rm d}\phi/{\rm d}t$) depends only on $\phi$ and temperature $T$,
i.e. only on the state variables and not on the temperature path by
which it had arrived to that state, then $\phi$ can be considered as
an additive quantity. This situation occurs if the nucleation sites
were consumed early in the reaction, i.e. site saturation occurs, and
if the growth rate is a function of instantaneous temperature only.

Cahn's additivity principle has been used by many workers for the
evaluation of non-isothermal transformations in materials; e.g.,
austenite to ferrite+pearlite transformations in steels
\cite{Hawbolt_Chau_Brimacombe_1983,Leblond_Devaux_1984,Umemoto_et_al_1983}
and a similar kind of transformation in titanium alloys under cooling
\cite{Damkroger_Edwards_1990,Malinov_et_al_2001}. A number of
investigators have evaluated the applicability of the additivity rule
in detail
\cite{Wierszy_1991,Lusk_Jou_1997,Zhu_et_al_1997,Reti_Felde_1999,Kampen_et_al_2002}.
A general modular approach, which accounts for the three phase
transformation mechanisms, nucleation, growth and impingement
applicable to both isothermal and isochronous conditions, have been
developed, discussed and applied to phase transitions in alloys
\cite{Mittemeijer_Sommer_2002,Liu_et_al_2004a,Liu_et_al_2004b}. A
recent overview of numerical and analytical methods for the
determination of the kinetic parameters of a phase transformation is
provided in \cite{Liu_et_al_2007}. An analytical solution for
non-isothermal KJMA rate equation comprising separate
activation energies for nucleation and growth when the transformation
occurs under continuous heating has been obtained
\cite{Farjas_Roura_2006}. In an integral concurrent approach, Elder et
al. \cite{Elder_et_al_1996} have studied the non-isothermal glassy or
amorphous metals (such as Fe-B, Cu-Zr and Mg-Zn alloys) by solving a
set of coupled physically-based equations for time-dependent spatially
uniform temperature field, for the nucleation and growth of single
crystallite and the Kolmogorov \cite{Kolmogorov_1937} formula for
$\phi$, simultaneously.

In this paper, we employ a method for calculation of the volume
fraction of the new phase as a function of time and temperature during
phase transformation in non-isothermal conditions. The method
satisfies Cahn's additivity rule and assumes that the system is not
too far away from equilibrium. It requires the specification of the
fraction of transformed phase in equilibrium as a function of
temperature. The method has been used to compute the phase
transformation behaviour of a zirconium alloy under both slow and
rapid heating and cooling (up to $\pm 100$ Ks$^{-1}$). It is
equivalent to the model for grain boundary nucleation when the
nucleation rate is high and site saturation occurs early during the
reaction \cite{Cahn_1956a}. The applicability of the method could be
computations of transformation behaviour during fabrications subject
to a variety of heat treatments, for example \cite{Massih_et_al_2003},
or in-service material performance under extreme conditions
\cite{Forgeron_et_al_2000}. We shall also outline a general method for
treating cases for which the volume fraction of the transformed phase
is thermal history dependent.

The organization of this paper is as follows. In section
\ref{sec:model}, a kinetic model for non-isothermal transformation, in
differential form and in integro-differential form, is presented.
The application of the differential method to the solid state phase
transformation in a zirconium alloy is presented in section
\ref{sec:apply}.  In section \ref{sec:discuss}, we discuss the
appropriateness of the model both from a theoretical stance and
empirical applicability. In section \ref{sec:summary}, we summarize
the main results and remark on possible future directions.

\section{Kinetic model}
\label{sec:model}

One common approach to model the kinetics of non-isothermal phase
transformation is to utilize the so called additivity rule
\cite{Avrami_1940,Cahn_1956b}. The rule may be stated as follows: The
temperature history (temperature vs. time) is subdivided into a number
of small type steps. Then the time spent for a volume element of the
material in time interval $\Delta t_i$ at a given temperature $T_i$
divided by the incubation time $t_{xi}$ at which the reaction has
attained a certain fraction $x$ of completion, can represent the
fraction of the total nucleation time required for formation of the
new phase. When the sum of such fractions reaches unity, the
transformation starts to occur.  Symbolically, it is expressed as
\begin{equation}
\sum_{i=1}^{n}\frac{\Delta t_i}{t_{xi}}=1 \quad \mbox{or} \quad \int_0^t\frac{{\rm d}s}{t_x[T(s)]}=1.
\label{eqn:scheil-rule}
\end{equation}
\noindent

A suitable material parameter for tracing the progress of solid-state
phase transformation is the transformed volume fraction $y$
as a function of  time $t$ and temperature $T$ with the property $0 \le y \le
1$.

Suppose the transformation is additive and the rate of
transformation  is only a function of the amount of
transformation and temperature \cite{Cahn_1956b}, namely

\begin{equation}
 \frac{{\rm d} y }{{\rm d}t} = f(y,T),
\label{eqn:y-rate}
\end{equation}
\noindent
The time it takes for a certain fraction $y_x$ of the new phase to be
completed is
\begin{equation}
 t_x = \int_0^{y_x} f(y,T)^{-1}dy,
\label{eqn:timex}
\end{equation}
\noindent
Equation (\ref{eqn:y-rate}) is a sufficient condition for the
additivity, since it satisfies the additivity rule
(\ref{eqn:scheil-rule}) through equation (\ref{eqn:timex}).

Following \cite{Leblond_Devaux_1984}, we consider that $y$ is
not too far from its steady-state or equilibrium value $y_s(T)$ at a
given temperature $T$, with $f(y_s,T)=0$. Thus, the first term in a
series expansion of $f$ about $y_s$ yields
\begin{equation}
 \frac{{\rm d} y }{{\rm d}t} = \frac{y_s(T)-y}{\tau_c(T)},
\label{eqn:lebdev-eq}
\end{equation}
\noindent
where $\tau_c$ is a characteristic time of phase transformation and
formally corresponds to $\tau_c=\big(\partial f/\partial
y|_{y=y_s}\big)^{-1}$.  Note that equation (\ref{eqn:lebdev-eq}) is
controlled by two external temperature-dependent functions, i.e.,
$\tau_c(T)$ and $y_s(T)$, with $y_s(T)$ being the fraction of a new
phase reached at temperature $T$ after infinitely long time and $0\le
y_s(T) \le 1$. Both these functions are material specific
temperature-dependent quantities and can be deduced from experimental
data on properties of a particular material or derived from
appropriate models verified with such data. By putting $\phi \equiv y/y_s$
and $k(T)\equiv\tau_c^{-1}(T)$, we rewrite equation (\ref{eqn:lebdev-eq})
in a reduced form
\begin{equation}
\frac{{\rm d} \phi }{{\rm d}t} = k(T)(1-\phi).
\label{eqn:lebdev-eq-norm}
\end{equation}
\noindent
Integrating this equation gives
\begin{equation}
\phi(t) = 1-\exp\Big(-\int_{t_0}^t k[T(s)]{\rm d}s\Big).
\label{eqn:lebdev-sol}
\end{equation}
\noindent

\begin{figure}
\begin{center}
\includegraphics[width=0.75\textwidth]{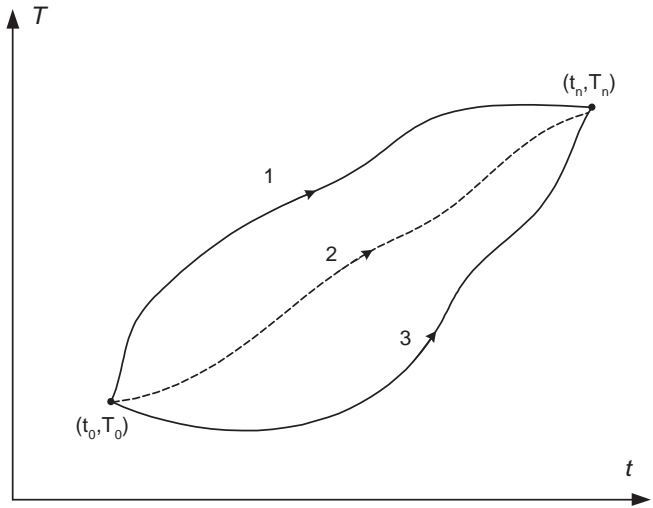}\\
\end{center}
\caption{Time ($t$) temperature ($T$) histories. A sample  undergoing  a
phase transformation when subjected to
a thermal history from state ($t_0,T_0$) to ($t_n,T_n$) via different
paths would reach different stages of transformations at ($t_n,T_n$).}
\label{fig:temp-path}
\end{figure}

 Under non-isothermal conditions (heating/cooling), $\phi$ not
only depends on $t$ and $T$ directly, but also on the way it has
reached that temperature at a given point in time, see figure
\ref{fig:temp-path}. This process may formally be described by means
of a generalized Langevin-type equation. On these conditions, the
temperature follows a path which may be an arbitrary function of time,
and hence equation (\ref{eqn:lebdev-eq}) needs to be solved
numerically (section \ref{sec:compute}).

We next present a  more general theoretical
treatment of the kinetics of phase transformation. As
pointed out in \cite{Cahn_1956b}, if the transformation rate
depends only on $y$ and $T$, i.e. on the state of the system rather on
the thermal path by which it has reached that state, then the
transformation is additive in the sense of equation
(\ref{eqn:scheil-rule}). By making the ansatz $f(\phi,T)=h(T)g(\phi)$,
we write equation (\ref{eqn:y-rate}) in the form
\begin{equation}
 \frac{{\rm d} \phi }{{\rm d}t} = h(T)g(\phi).
\label{eqn:phi-rate2}
\end{equation}
\noindent
As an example, for the KJMA model (\ref{sec:kjma}), we write
\begin{eqnarray}
h(T) & = & m k(T)^{1-\gamma}\\
g(\phi) & = & [-\ln(1-\phi)]^\gamma (1-\phi),
\label{eqn:kjma-vars}
\end{eqnarray}
\noindent
with $\gamma=(m-1)/m$ and $m$ is the overall growth exponent
(\ref{sec:kjma}). In our special case, $m=1$, thereby
(\ref{eqn:phi-rate2}) yields equation (\ref{eqn:lebdev-eq}).

We generalize the differential equation (\ref{eqn:phi-rate2}) to
overcome the restriction of additivity \cite{Hermansson_2008}. In the
manner of \cite{Mori_1965a}, we write
\begin{equation}
 \frac{{\rm d} \phi }{{\rm d}t} = \int_0^tM[T(t)-T[\bar{t})]g[\phi(\bar{t})]{\rm
d}\bar{t}+\eta(t),
\label{eqn:mori-eq}
\end{equation}
\noindent
where $M(t)$ is a memory kernel, which accounts for the effect of
non-additivity and $\eta(t)$ is a random force, e.g. a Gaussian white
noise, which stands for the thermal fluctuations of the system during
phase transformation. An important property of $\eta(t)$ is that it
vanishes on the average, i.e. $\langle\eta(t)\rangle=0$, and it is
un-correlated to $\phi(t)$, i.e. $\langle\phi(t)\eta(t)\rangle=0$, see
e.g. \cite{Forster_1975}. Also, $M(t)$ is proportional to the spectrum
of the random force, viz.
\begin{equation}
 M(t-t^\prime)\propto \frac{1}{k_BT} \langle\eta(t)\eta(t^\prime)\rangle .
\label{eqn:memo-fun}
\end{equation}
\noindent
If now $M(t)$ decays in a time $\tau_M$ and the applicable time is much
longer than  $\tau_M$, then
\begin{equation}
M(T[t]-T[\bar{t})])= h(T)\delta (t-\bar{t}),
\label{eqn:memo-fun2}
\end{equation}
\noindent
where $\delta(t)$ is the Dirac delta function. Thus, with relation
(\ref{eqn:memo-fun2}), Eq. (\ref{eqn:mori-eq}) reduces to
(\ref{eqn:phi-rate2}), that is when only the additivity rule is applicable.

Let us now consider the case of non-isothermal transformation
with a path-dependent characteristic. Suppose the volume fraction
transformed obeys a general relation in the form
\begin{equation}
\frac{{\rm D} y}{{\rm D}T} = F[\varphi(T)],
\label{eqn:y-functional}
\end{equation}
\noindent
where ${\rm D}y/{\rm D}T$ is the functional derivative of $y$ with
respect to $T$ and $F[\varphi(T)]$ is a functional of $\varphi$ in the
sense that not only it depends on particular values of $T$ and $t$,
but on the function $\varphi$ and all of $T$ and $t$ which are covered
by $\varphi$, moreover $0\le F[\varphi(T)] \le 1$. The form of this
functional defines the transformation rule. The function $\varphi$ may
be described as a time integral of a rate of transformation $k(T)$
\cite{Cahn_1956b,Mittemeijer_1992}
\begin{equation}
\varphi[T(t)] = \int_0^tk(T(s)){\rm d}s.
\label{eqn:phi-function}
\end{equation}
\noindent

In case of interest to evaluate all the possible paths in the phase
transformation from point $(T_0,t_0)$ to $(T,t)$ in the
temperature-time plane, we re-express equation
(\ref{eqn:y-functional}) in terms of a path-integral in the form
\begin{equation}
y(T,t;T_0,t_0)= \int_{T_0,t_0}^{T,t} F[T(s),\dot{T}(s)]{\rm d}[T(s)],
\label{eqn:y-path}
\end{equation}
\noindent
where $\dot{T}={\rm d}T/\rm dt$ and ${\rm d}[T(t)]$ denotes the product of
infinitesimal steps in the thermal history, i.e.,
${\rm d}[T(t)]=\prod_{j=1}^n{\rm d}T_j$, with ${\rm d}T_j$ being the
temperature increment at time $t_j$. Hence, in the path-integral
formulation, the time interval $(t_0,t)$ is
divided into $n+1$ intervals of equal length separated by time points
$t_1, t_2, \dots t_n,t$ at which a material volume element (particle)
is at temperatures $T_1, T_2, \dots T_n,T$, respectively. Assigning a
functional form for $F$ by using a suitable model for isothermal phase
transformation, a path-dependent description for the evolution of $y$
under non-isothermal conditions is obtained.

We choose a functional form for $F[T(t),\dot{T}(t)]$
in equation (\ref{eqn:y-path}) as
\begin{equation}
  F[T(t),\dot{T}(t)]=\delta[T(t)]-\exp\Big\lbrace-\Big[\int_{t_0}^tk[T(s),\dot{T}(s)]{\rm d}s\Big]^m\Big\rbrace
  \label{eq:prob-dens}
\end{equation}
\noindent
where $\delta[T(t)]$ is a functional Dirac delta distribution. Substituting equation
(\ref{eq:prob-dens}) into (\ref{eqn:y-path}), with
normalization, gives
\begin{equation}
  \phi(T,t)=1-\int_{T_0,t_0}^{T,t}\exp\Big\lbrace-\Big[\int_{t_0}^tk[T(s),\dot{T}(s)]{\rm d}s\Big]^m\Big\rbrace
 {\rm d}[T(t)].
  \label{eq:kjma-propagator}
\end{equation}
\noindent
This equation represents a path-integral description (or propagator) of the KJMA
relation for the new phase development in non-isothermal conditions;
and the functional integrand (\ref{eq:prob-dens}) may be interpreted
as the probability-density for the new phase to follow a specific
trajectory $T(t)$, see figure \ref{fig:temp-path}.

\section{Application}
\label{sec:apply}
\subsection{Experimental data on zirconium alloys}
\label{sec:experiment}
In this subsection, we make a short survey of experiments reported in literature, which
we have used to select the input model parameters for the application
of our model to zirconium base alloys. These experiments also provide
data for model retrodictions and predictions. We consider the kinetics
of phase transformation of Zircaloy-4 (Zr-1.5Sn-0.2Fe-0.1Cr-0.12O, by
wt\%). Zirconium in solid state undergoes an allotropic transformation
from the low temperature hexagonal closed-packed (hcp) $\alpha$-phase
to body-centered cubic (bcc) $\beta$-phase at 1138 K
\cite{Lemaignan_Motta_1994}. On cooling, the transformation is either
bainitic or martensitic depending on the cooling rate, with a strong
epitaxy of the $\alpha$-platelets in the former $\beta$ grains
\cite{Lemaignan_Motta_1994}.

Solid state phase equilibria of Zircaloy-4 have been investigated
experimentally  \cite{Miquet_et_al_1982a}, who reported a
prevalence of four phase domains: $\alpha+\chi$ up to 1081 K,
$\alpha+\beta+\chi$ from 1081 to 1118 K, $\alpha+\beta$ between 1118
and 1281 K, and $\beta$-phase above 1118 K. Here, $\chi$ refers to the
intermetallic hexagonal Laves phase Zr(Fe,Cr)$_2$, see e.g.
\cite{Bangaru_et_al_1987}. Quenching Zircaloy from $\beta$-phase in
moderate cooling rates produces two variants of Widmanst\"{a}tten
structure, namely, the basketweave and the parallel-plate structure
\cite{Okvist_Kallstrom_1970,Massih_et_al_2003}. However, at cooling
rates greater than 1000 Ks$^{-1}$ a martensite structure is
observed, while for very slow cooling rates, $\le 0.5$ Ks$^{-1}$, the
needle-shaped structure is rarely seen
\cite{Charquet_Alheritiere_1987}.  The overall $\alpha \leftrightarrow
\beta$ transition in Zircaloy-4 has been studied by a number of
workers in the past
\cite{Holt_1970,Holt_1973,Woo_Tangri_1979,Hunt_Schulson_1980,Corchia_Righini_1981,Yoo_Kim_1991}
and more recently \cite{Forgeron_et_al_2000,Brachet_et_al_2002}, which
include also experiments on Zr-Nb alloys.

Forgeron et al. \cite{Forgeron_et_al_2000} studied $\alpha \leftrightarrow
\beta$ transition of the Zr alloys by determining both their equilibrium
(steady-state) temperature-dependence and their transient behaviour,
with respect to the fraction of volume transformed, for the heating/cooling
rates from $\pm$0.1 to $\pm$100 Ks$^{-1}$. More specifically,
Forgeron et al. \cite{Forgeron_et_al_2000} determined the equilibrium behaviour
of $\alpha/\beta$-phase fraction as a function of temperature by means
of calorimetry measurements. They used slow heating/cooling rates from
0.1 K/minute to 20 K/minute. Moreover, they carried out direct
measurements of the $\alpha/\beta$-phase fraction by employing image
analysis techniques on samples annealed for a few hours at different
temperatures then quenched to room temperature. For kinetic
measurements, they used dilatometry, where thermal
cycles were applied on tubular samples, 12 mm in length, in vacuum or
helium gas. As for calorimetric measurements, the uncertainty of the
relative phase fraction measurement was less than 5\%.

\subsection{Computations}
\label{sec:compute}
In the model for computation of the relative phase fraction as a
function of time and temperature, two functions $y_s(T)$ and
$\tau_c(T)$ appear in equation (\ref{eqn:lebdev-eq}), which need to be
specified. Let us consider first the former. The experimental data for
the temperature dependence of steady-state volume fraction under
phase transition suggest that $y_s(T)$ has an S-shaped or sigmoid
form. For this reason, we have selected for the $\alpha \leftrightarrow
\beta$ transition in Zr alloys a relation of the form
\begin{equation}
 y_s = \frac{1}{2}\Big[1-\tanh\Big(\frac{T-T_{cent}}{T_{span}}\Big)\Big],
\label{eqn:ss-sol}
\end{equation}
\noindent
for the equilibrium $\beta$-phase volume fraction at temperature
$T$. Here, $T_{cent}$ and $T_{span}$ are material specific parameters
that are related to the center and the span of the mixed-phase temperature
region, respectively. They are determined from the measured phase boundary
temperatures $T_{\alpha}$ and $T_{\beta}$ through
\begin{equation}
 T_{cent} = \frac{T_\alpha+T_\beta}{2}; \qquad  T_{span} = \frac{T_\beta-T_{cent}}{2.3}.
\label{eqn:T-phase-boundary}
\end{equation}
\noindent
Here, $T_{\alpha}$ and $T_{\beta}$ are defined as the temperatures
that correspond to 99\% $\alpha$- and  $\beta$-phase fractions,
respectively. For Zircaloy-4, the data in
\cite{Miquet_et_al_1982b} gives
$T_{\alpha}=1079$ K and $T_{\beta}=1273$ K. We have used the data in
\cite{Forgeron_et_al_2000,Brachet_et_al_2002} to obtain $T_{cent} =
1159$ and $T_{span} = 44$ K.

\begin{figure}
\begin{center}
\includegraphics[width=0.80\textwidth]{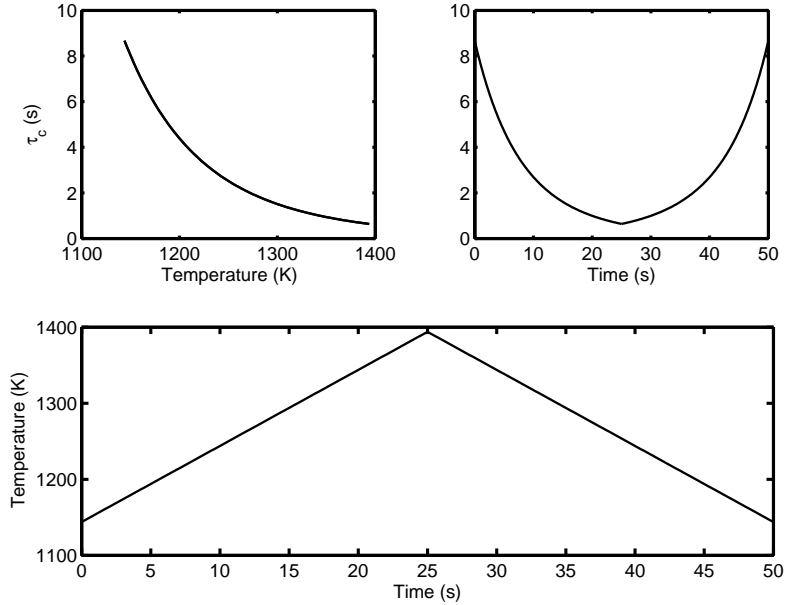}\\
\end{center}
\caption{Temperature dependence of the characteristic time for
Zircaloy-4 in the range of interest (top left) and its time-dependence
for a triangular-shaped temperature pulse (bottom panel).}
\label{fig:tau-gen}
\end{figure}

\begin{figure}
\begin{center}
\includegraphics[width=0.80\textwidth]{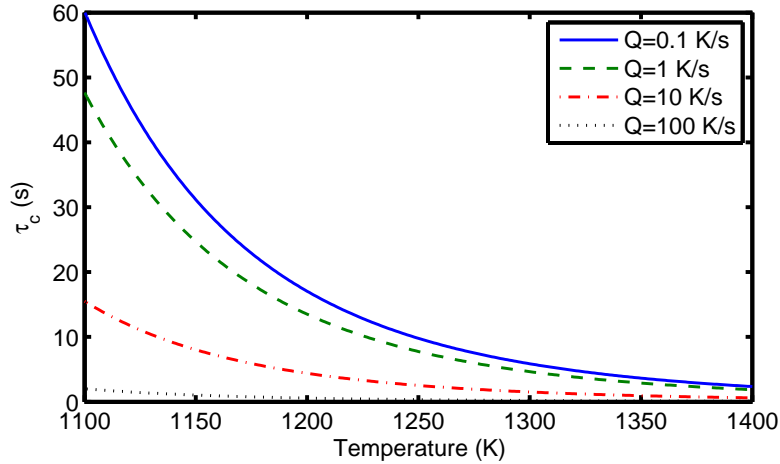}\\
\end{center}
\caption{Characteristic time as a function of temperature and
temperature rate for Zircaloy-4 according to
equation (\ref{eqn:rate-para}).}
\label{fig:tauq-gen}
\end{figure}

The relation for $\tau_c(T)$ or its inverse the rate parameter, in general,
$k=\tau_c^{-1}$ depends on both the nucleation rate and growth rate of
the new phase, which are strongly temperature dependent. It is usual
to adopt an Arrhenius-type relation for the rate parameter
\cite{Mittemeijer_1992,Kampen_et_al_2002}
\begin{equation}
 k(T) = k_0\exp\Big[-\frac{E}{k_BT(t)}\Big],
\label{eqn:rate-para}
\end{equation}
\noindent
where $k_0$ is a kinetic prefactor, $E$ the overall effective
activation energy  and $k_B$ the Boltzmann
constant. For Zircaloy-4, using the
data on volume fraction \cite{Forgeron_et_al_2000}, we found best fit values:
$k_0=60457+18129|Q|$ (s$^{-1}$) and $E/k_B=16650$ (K), where $Q={\rm
d}T/{\rm d}t$ is the heat rate (Ks$^{-1}$) in the range $0.1 \le |Q|
\le 100$ Ks$^{-1}$. Typical variations of  $\tau_c(T)=k(T)^{-1}$ for a triangular
temperature pulse is shown in figure \ref{fig:tau-gen}. Figure
\ref{fig:tauq-gen} illustrates the  temperature rate
dependence of $\tau_c$  as a function of temperature according to
equation (\ref{eqn:rate-para}).

The experimental results on Zircaloy-4 indicate that the starting
temperature for the onset of phase transformation is temperature rate
dependent \cite{Forgeron_et_al_2000}. This phenomenon has also been
observed in other systems, e.g., ferritic-pearlitic transformation in
steels \cite{Leblond_Devaux_1984,Umemoto_et_al_1983} and $\alpha \leftrightarrow
\alpha+\beta$ transition of the titanium alloys
\cite{Damkroger_Edwards_1990}. In our modeling of this effect, for
Zircaloy-4, we have related the onset of $\alpha \to \alpha+\beta$
transformation temperature (heating) by
fitting a power law relation to the experimental data reported in
\cite{Forgeron_et_al_2000,Brachet_et_al_2002}, see appendix A of \cite{Massih_2009}.

\begin{figure}
\begin{center}
\includegraphics[width=0.80\textwidth]{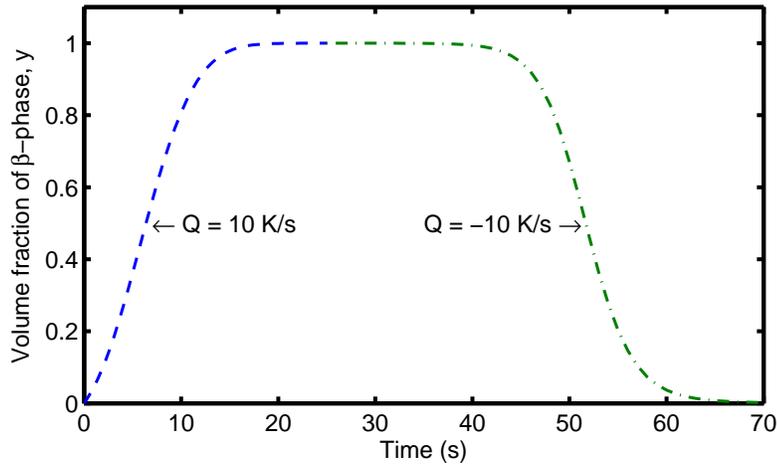}\\
\end{center}
\caption{Calculated volume fraction of $\beta$-phase in Zircaloy-4
versus time during heating and subsequent cooling at the rates of
$\pm$10 K/s, using an extended version of the temperature history shown
in figure \ref{fig:tau-gen}.}
\label{fig:y-tid}
\end{figure}

\begin{figure}
\begin{center}
\includegraphics[width=0.80\textwidth]{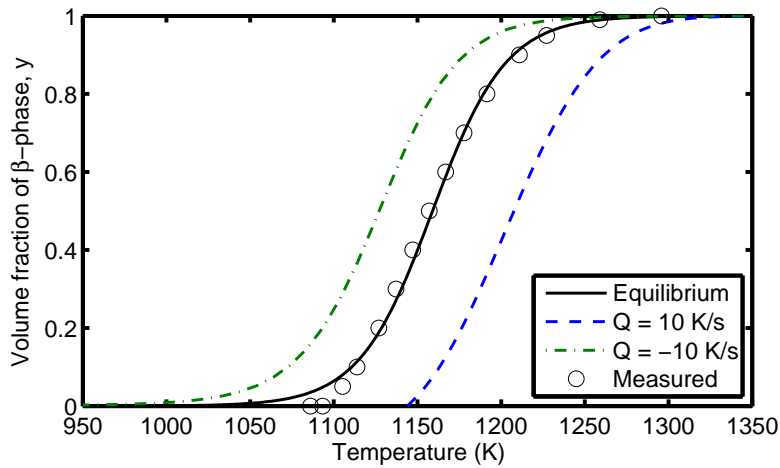}\\
\end{center}
\caption{Calculated volume fraction of $\beta$-phase in Zircaloy-4
versus temperature during heating and subsequent cooling at the rates of
$\pm$10 K/s, using an extended version of the temperature history shown
in figure \ref{fig:tau-gen}. The equilibrium curve is calculated using
equation (\ref{eqn:ss-sol}) and the corresponding measured data are
from Forgeron et al. (2000).}
\label{fig:y-temp}
\end{figure}

As an example, for simulation of non-isothermal phase transformation,
we have calculated the fraction of volume transformed during $\alpha
\leftrightarrow \beta$ transition of Zircaloy-4 material by solving
equation (\ref{eqn:lebdev-eq}) employing a similar triangular
temperature history shown in figure \ref{fig:tau-gen}, but extending
its tail to 75 s.  The initial condition utilized for heating is
$y(0)=0$, while on cooling $y(t_f)=1$, where $t_f$ corresponds to the end time
of heating or the beginning of cooling. The Runge-Kutta method of order 4 and 5
\cite{Quarteroni_Saleri_2003} was used to integrate equation
(\ref{eqn:lebdev-eq}). Figure \ref{fig:y-tid} shows $y$ as a function
of time and figure \ref{fig:y-temp} depicts $y$ versus temperature. In
figure \ref{fig:y-temp}, we have included the equilibrium curve using
equation (\ref{eqn:ss-sol}) and the corresponding experimental data
reported in \cite{Forgeron_et_al_2000}. The computations on
heating/cooling at the rates $\pm 10$ K/s are in agreement with the
data reported in \cite{Forgeron_et_al_2000}. We note the
asymmetry between the heating and the cooling in the
computations. This is a reflection of experimental data from which the model
has been adjusted to, which in turn emanates from the thermodynamic
instability of the system. A more detailed comparison between experimental
data and model computations, at several heating/cooling rates for
zirconium alloys, is presented in \cite{Massih_2009}.

\section{Discussion}
\label{sec:discuss}
The method utilized here to calculate the volume fraction of the
favoured phase as a function of time and temperature can be applied to
any input thermal history since we solve equation
(\ref{eqn:lebdev-eq}) numerically. The question that may arise is how
well this approximate equation compares with the prevailing exactly
solved models in isothermal conditions.

For isothermal conditions, equation (\ref{eqn:lebdev-sol}) yields
\begin{equation}
 \phi = 1-e^{-k(T)(t-t_0)},
\label{eqn:lebdev-isosol}
\end{equation}
\noindent
where $t_0$ is the incubation time for the onset of phase
transformation. This relation is a special case of the Avrami model
\cite{Avrami_1940,Cahn_1956a} related to site saturation
transformation on grain surfaces under isothermal conditions, namely
\begin{equation}
 \phi = 1-\exp\big[-\frac{2S}{V}v(t-t_0)\big],
\label{eqn:avrami}
\end{equation}
\noindent
where $S/V$ is the surface area to volume ratio and $v$ the interface
velocity (growth rate) of the nucleus. Moreover, as has been shown by
Cahn \cite{Cahn_1956a}, equation (\ref{eqn:avrami}) is theoretically
exact for the systems in which the nucleation rate is sufficiently
rapid such that site saturation occurs early in the reaction
period. Indeed, he calculated that the time at which the site
saturation occurs is $t_s=(I_s v^2)^{-1/3}$, where $I_s$ is the
nucleation rate per unit area of the grain boundary, see
 \ref{sec:kjma}.  Our considered model is related to a special case of
the KJMA theory where nucleation rate is high and site saturation
occurs early in reaction, see  \ref{sec:kjma}.

\begin{figure}
\begin{center}
\includegraphics[width=0.80\textwidth]{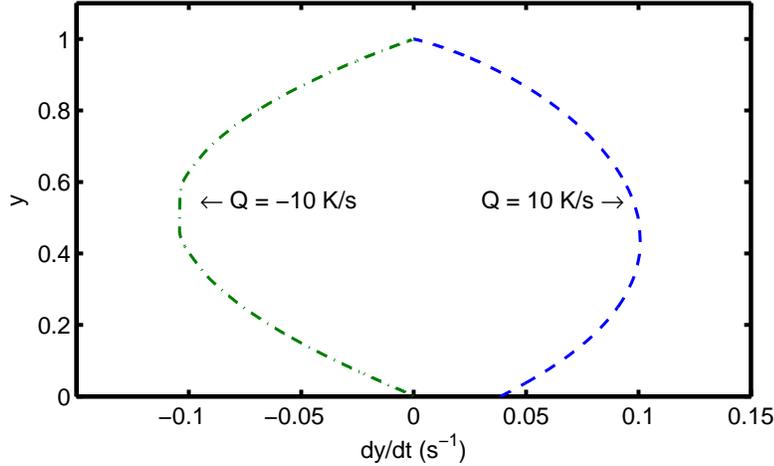}\\
\end{center}
\caption{Calculated volume fraction $y$ of $\beta$-phase in Zircaloy-4
versus its rate $\rm dy/\rm dt$ during heating and subsequent cooling
at the thermal rates of
$\pm$10 K/s, using an extended version of the temperature history shown
in figure \ref{fig:tau-gen}.}
\label{fig:y-portrait}
\end{figure}

A noteworthy observation in our computations, which is a reflection of
experiments reported in \cite{Forgeron_et_al_2000}, is the
asymmetry in phase transformation between heating and cooling carried
out at the same thermal rating, cf. figures \ref{fig:y-tid} and
\ref{fig:y-temp}. To illustrate this asymmetry, we have plotted the
phase portrait of our computations in figure
\ref{fig:y-portrait}. This is a manifestation of the different initial
conditions and the different paths of the two processes. In our case,
we had to extend the cooling tail of the temperature history
(cf. figure \ref{fig:tau-gen}) to 950 K to complete the phase
transformation, see figure \ref{fig:y-temp}. A thermodynamic
explanation of this phenomenon even at a phenomenological level is
hard for us to ascertain here; our model is purely a kinetic one.

\begin{figure}
\begin{center}
\includegraphics[width=0.80\textwidth]{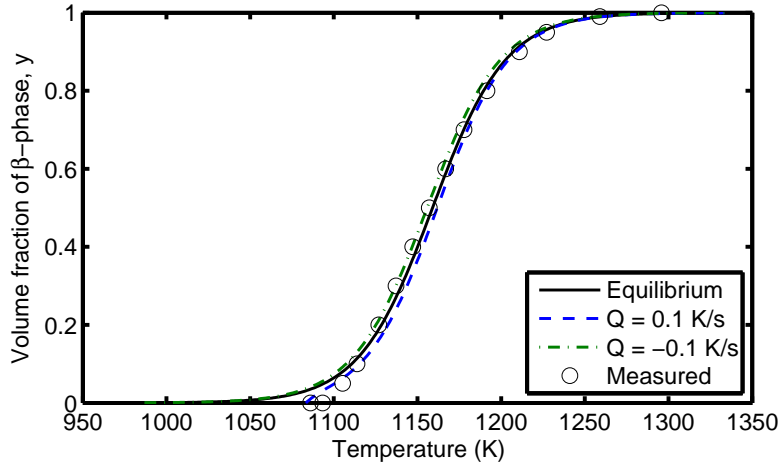}\\
\end{center}
\caption{Calculated volume fraction of $\beta$-phase in Zircaloy-4
versus temperature during heating and subsequent cooling at the rates of
$\pm 0.1$ Ks$^{-1}$, using an expanded version of the temperature history shown
in figure \ref{fig:tau-gen}. This figure should be compared with figure
\ref{fig:y-temp}.}
\label{fig:y-temp-equilb}
\end{figure}

The asymmetry is also true for the thermal rate dependence of the
start temperature of phase transformation, cf. equations in appendix A
of \cite{Massih_2009}. This type of hysteresis is ubiquitous in
first-order phase transitions in solids.  It is a manifestation of the
solid state supercooling effect observed, e.g. in zirconium alloys,
even at slow coolings
\cite{Miquet_et_al_1982,Canay_Danon_Arias_2000}. One reason for this
behaviour is that the transformation process under non-isothermal
conditions is away from equilibrium and is accompanied by dissipation
of energy. The hysteresis should decrease with the decrease in
transformation rate and should disappear at infinitely slow
transitions. In other words, at infinitely low heating/cooling rates,
the temperature variation of $y$ should collapse on the equilibrium
curve. To check this, we have repeated the aforementioned computations
with a thermal rate two orders of magnitude lower, i.e. at $\pm 0.1$
Ks$^{-1}$ by expanding the thermal history over 6000 s. The results
are depicted in figure \ref{fig:y-temp-equilb}, which should be
compared with figure \ref{fig:y-temp}. We should, nevertheless,
mention that in literature, the start temperature of first order phase
transition has been related to the temperature dependence of the
incubation time for nucleation \cite{Zhu_Devletian_1991,Massih_2009}.

\section{Summary and outlook}
\label{sec:summary}
In summary, we have studied the overall phase transformation kinetics
in alloys under non-isothermal conditions using a differential analysis
method. The method was applied successfully to the massive phase
transformation in a zirconium alloy, for which the time/temperature
variations of the volume fraction of the favoured phase under heating
and cooling were calculated. A trait of non-isothermal or
non-equilibrium phase transitions is the dependence of the onset of
the transformation temperature on heating/cooling rates. We used
best-fit empirical relations to account for this effect for
Zircaloy. However, we alluded to the intimate relationship between
this temperature and the incubation time for nucleation. This
relationship warrants further analysis and may be studied in the
context of transient nucleation phenomenon.

Non-isothermal transformation kinetics is usually treated with the
additivity rule, where the transient heating/cooling is treated as a
series of small isothermal steps. A formulation, which embraces both
additive and non-additive situations, has been suggested. Moreover, a
method based on the concept of path-integral, which accounts for all
the possible thermal histories to reach the final state, has been
presented. The path-integral approach has the potential to provide the
optimal thermal path for phase transformation to attain the desired
microstructure, and thereby material's macroscopic behavior.

\paragraph{Acknowledgments}
We are indebted to Patrik Hermansson for valuable communications and
ARM thanks Richard Warren for discussions. The work was
supported in part by the Swedish Radiation Safety Authority under the contract
number SKI2007/611, 200806001.

\appendix
\section{Kolmogorov-Johnson-Mehl-Avrami model}
\label{sec:kjma}
The overall phase transformation kinetics describes the
time/temperature evolution of the volume fraction of the newly
transformed phase onto the stable phase. The transformations are
supposed to occur by nucleation and growth mechanisms (i.e. first order
transition). A mean field model was formulated and solved exactly by
Kolmogorov \cite{Kolmogorov_1937},
Johnson and  Mehl \cite{Johnson_Mehl_1939}, Avrami \cite{Avrami_1939},
Evans \cite{Evans_1945}, and Jackson \cite{Jackson_1974}, apparently all
independently.  The basic assumptions of the model are : (i) the
system is infinite in extent, and hence, boundary effects are neglected;
(ii) nucleation is a stochastic process and on average takes place uniformly;
and (iii) the growth of new phase grains ceases at the mutual points
of their contact, but continues vigorously elsewhere; see
ref. \cite{Burbelko_et_al_2005} for a recent appraisal.

Kolmogorov \cite{Kolmogorov_1937} rigorously derived a relation for the
time evolution of the transformed phase $\phi$ in three
dimensions. In a $d$-dimensional space,  we write

\begin{equation}
 \phi(t) = 1-e^{-\phi_e(t)},
\label{eqn:kjma_vol}
\end{equation}
\noindent
where $\phi_e(t)$ is the so-called extended volume fraction, expressed
as
\begin{equation}
 \phi_e(t) = C_d\int_0^tI(s)R^d(t,s){\rm d}s,
\label{eqn:kjma_exvol}
\end{equation}
\noindent
where $I(t)$ is the nucleation rate (per unit volume and time),
$R(t,s)$ is the radius of the new grain (phase) at time $t$ that was
nucleated at time $s$, and $C_d$ is a shape factor given for a
hypersphere by
\begin{equation}
 C_d = \frac{2\pi^{d/2}}{d\,\Gamma(d/2)},
\label{eqn:geofact}
\end{equation}
\noindent
with $\Gamma(\bullet)$ denoting the usual gamma function. The function
$\phi_e(t)$ is the sum of volumes of all the transformed grains
divided by the total volume of the system, supposing that the grains
never cease growing and the new ones keep nucleating at the same rate
in the entire material.

For $R>>R^\ast$, where $R^\ast$ is the critical radius for nucleation,
$R(t,s)$ can be expressed in terms of the growth rate of the new phase as
\begin{equation}
 R(t,s) = \int_{s}^t v(\tau)\rm d\tau,
\label{eqn:velocity-integ}
\end{equation}
\noindent
where $v$ is the interfacial growth velocity. If now, the nucleation rate
and the interfacial velocity assume anomalous power law dependencies
on time, namely, $I(t)\simeq I_0t^\alpha$ and $v(t)\simeq v_0t^\beta$,
then using equation (\ref{eqn:velocity-integ}), equation
(\ref{eqn:kjma_exvol}) can be  expressed as
\begin{equation}
 \phi_e(t) = \Big(\frac{t-t_{inc}}{\tau_g}\Big)^m,
\label{eqn:kjma_exvol_anom}
\end{equation}
where $m$ is an overall growth exponent of the favoured phase
\noindent
\begin{equation}
 m = 1+\alpha+(1+\beta)d,
\label{eqn:kjma_exponent}
\end{equation}
\noindent
and $\tau_g$ is a characteristic time for the growth of the favoured phase
\begin{equation}
 \tau_g = \Bigg[\frac{(1+\beta)^{d+1}}{C_dI_0v_0^d\Gamma(1+d)}\Big(\frac{1+\alpha}{1+\beta}\Big)_{1+d}\Bigg]^{1/m},
\label{eqn:kjma_tau}
\end{equation}
\noindent
with $(z)_n=\Gamma(z+n)/\Gamma(z)$ being a Pochhammer symbol.

Here, we have shifted the time origin as $t \Rightarrow t-t_{inc}$,
according to modelling of transient nucleation in
\cite{Shneidman_Weinberg_1993}, where $t_{inc}$ is called the
incubation time and is found to be
\begin{equation}
 t_{inc} = t^\ast\ln\frac{t}{t^\ast}\frac{\Phi^\ast}{T},
\label{eqn:incub-time}
\end{equation}
\noindent
where $t^\ast=R^\ast/v_0$ and $\Phi^\ast$ is the energy barrier for
nucleation in kelvin; see also Iwamatsu's calculations for $d=2$ \cite{Iwamatsu_2008}. To make simplifications, recall the classical nucleation theory and assume that $I_0$ is the
steady-state nucleation rate with the form
$I_0=I^\ast\exp(-\Phi^\ast/T)$, i.e. the number of critical nuclei per
unit volume and time; e.g. see chapter 10 in
\cite{Christian_2002}. Hence, according to equation
(\ref{eqn:kjma_tau}), $\tau_g =
\tau^\ast\exp(\Phi^\ast/mT)$. Putting $t\approx\tau_g$ in
equation (\ref{eqn:incub-time}) gives
\begin{equation}
 t_{inc}  =
t^\ast\frac{\Phi^\ast}{mT}+t^\ast\Big[\ln\frac{\Phi^\ast}{T}-\ln\frac{t^\ast}{\tau^\ast}\Big]
\approx  t^\ast\frac{\Phi^\ast}{mT}.
\label{eqn:incub-time-app}
\end{equation}
\noindent
So, as $T$ is lowered, both $\tau_g$ and $t_{inc}$ are increased,
albeit their unlike temperature dependence. We note that in the
aforementioned formalism with $\alpha=0$, $\beta=0$ and $d=3$, with $m=4$, the
frequently used expression for the KJMA model is recovered.

Cahn \cite{Cahn_1956a} derived relations for $\phi_e$ for
particular cases where the new phase in a polycrystalline material
nucleates on grain boundary surfaces, grain edges, or grain corners,
under the condition that it grows with constant
velocity. Two limiting cases corresponding to high and low
nucleation rates (relative to growth rates) were found. At high
temperatures, the nucleation rate is low and site saturation may not
occur. If again $I \simeq I_0t^\alpha$ and assuming random nucleation
on either grain surfaces, edges, corners, or interiors, Cahn found
(for $d=3$)
\begin{equation}
 \phi_e  = 8\pi \frac{\Gamma(1+\alpha)}{\Gamma(2+\alpha+d)}I_0v_0^d t^m.
\label{eqn:phi-lowrate}
\end{equation}
\noindent
Moreover, since $I_0$ would be proportional to the number of available
nucleation sites, then $I_0 \propto 1/G^{d-j}$, where $G$ is the
parent phase grain size, $j=d-3$ for corner, $j=d-2$ for edge, $j=d-1$
for surface, and $j=d$ for intragranular nucleation. Equation
(\ref{eqn:phi-lowrate}), except for a numerical factor and a shift in
time origin, is close to the general equation
(\ref{eqn:kjma_exvol_anom}).

 For high nucleation rates, site saturation occurs early in the
reaction and $\phi_e$ becomes independent of nucleation rate. The
reaction is completed when $t=0.5G/v$ \cite{Cahn_1956a}
\begin{equation}
 \phi_e  = C_n\rho_n(vt)^n.
\label{eqn:phi-highrate}
\end{equation}
\noindent
Here, $n=1$, 2, 3 correspond respectively to site saturation on grain
surfaces, edges and corners; and $\rho_n$ denotes the ratio of the
respective surface area, grain-edge length and grain corner number per
unit volume; note that $ \phi_e$ is independent of nucleation rate
and is driven by the growth rate. If we put $v(t)\simeq v_0t^\beta$,
then $\phi_e  \simeq C_n\rho_nv_0^n t^{n(1+\beta)}\equiv
k_nt^{n(1+\beta)}$.

Thus, for $n=1$ (grain boundary site saturation) and $\beta=0$ equation
(\ref{eqn:kjma_vol}) can be written as
\begin{equation}
 \phi = 1-e^{-k_1t},
\label{eqn:kjma_n=1}
\end{equation}
\noindent
where $k_1=C_1\rho_1v_0=2(S/V)v_0$, which except a shift in the time
origin, is exactly the same as equations
(\ref{eqn:lebdev-isosol})-(\ref{eqn:avrami}), and is equivalent to the
model we have used in our analysis.

It is worthwhile to mention that Ham \cite{Ham_1959} obtained a
similar kind of relation as (\ref{eqn:kjma_n=1}) for diffusion-limited
precipitation of second phase in a crystalline. In Ham's model, one may
imagine that a polycrystalline consists of an array of parallel
cylinders, in which the solute atoms precipitate on cylinders'
surfaces. For this model, the precipitated fraction of excess solute
$w$ can be calculated as a function of time $t$ fairly accurately by
the simple relationship
\begin{equation}
  \label{eq:Hams_fraction_app}
  w \cong 1-e^{-t/\tau_0},
\end{equation}
\noindent
where $\tau_0=\ell^2/(\alpha_0^2D)$, $\ell$ is the inter-precipitate distance, $\alpha_0$ the lowest eigenvalue of the prevailing diffusion equation, and $D$ is the solute diffusivity. Hence, the theoretical basis of the model used for calculations of phase transformation kinetics in this paper may be attributed to a special case of the KJMA formulation, or diffusion-limited phase transformation on surfaces.

\section*{References}
\bibliographystyle{unsrt} \bibliography{kinetic}

\begin{thebibliography}{10}

\bibitem{Gunton_et_al_1983}
J.~D. Gunton, M.~San Miguel, and P.~S. Sahni.
\newblock The dynamics of first-order phase transition.
\newblock In C.~Domb and J.~L. Lebowitz, editors, {\em Phase Transitions and
  Critical Phenomena}, volume~8, chapter~3, pages 267--479. Academic Press,
  London, England, 1983.

\bibitem{Langer_1992}
J.~S. Langer.
\newblock In C.~Godr\`{e}che, editor, {\em Solids Far from Equilibrium},
  chapter~3, pages 297--363. Cambridge University Press, Cambridge, UK, 1992.

\bibitem{Porter_Easterling_1981}
D.A. Porter and K.E. Easterling.
\newblock {\em Phase Transformations in Metals and Alloys}.
\newblock Chapman $\&$ Hall, London, UK, 1981.
\newblock Chapter 5.

\bibitem{Kolmogorov_1937}
A.~N. Kolmogorov.
\newblock Statistical theory of metal crystallization.
\newblock {\em Izv. Akad. Nauk SSSR}, 1:355--359, 1937.
\newblock in Russian.

\bibitem{Johnson_Mehl_1939}
W.~A. Johnson and R.~F. Mehl.
\newblock Reaction kinetics in processes of nucleation and growth.
\newblock {\em Trans. of AIME}, 135:416--442, 1939.

\bibitem{Avrami_1939}
M.~Avrami.
\newblock Kinetics of phase change \mbox{I}: General theory.
\newblock {\em J. Chem. Phys.}, 7:1103--1112, 1939.

\bibitem{Avrami_1940}
M.~Avrami.
\newblock Kinetics of phase change \mbox{II}: Transformation-time relations for
  random distribution of nuclei.
\newblock {\em J. Chem. Phys.}, 8:212--224, 1940.

\bibitem{Cahn_1956b}
J.~W. Cahn.
\newblock Transformation kinetics during continuous cooling.
\newblock {\em Acta Met.}, 4:572--575, 1956.

\bibitem{Hawbolt_Chau_Brimacombe_1983}
E.~B. Hawbolt, B.~Chau, and J.~K. Brimacombe.
\newblock Kinetics of austenite-pearlite transformation in eutectoid carbon
  steel.
\newblock {\em Metall. Trans. A}, 14A:1803--1815, 1983.

\bibitem{Leblond_Devaux_1984}
J.~B. Leblond and J.~Devaux.
\newblock A new kinetic model for anisothermal metallurgical transformations in
  steels including the effect of austenite grain size.
\newblock {\em Acta Met.}, 32:137--146, 1984.

\bibitem{Umemoto_et_al_1983}
M.~Umemoto, K.~Horiuchi, and I.~Tamura.
\newblock Pearlite transformation during continuous cooling and its relation to
  isothermal transformation.
\newblock {\em Trans. ISIJ}, 23:690--695, 1983.

\bibitem{Damkroger_Edwards_1990}
B.~K. Damkroger and G.~R. Edwards.
\newblock Continuous cooling transformation kinetics in alpha-beta titanium
  alloys.
\newblock In M.P. Anderson, editor, {\em Simulation and Theory of Evolving
  Microstructures}, pages 129--150. The Minerals, Metals $\&$ Material Society,
  1990.

\bibitem{Malinov_et_al_2001}
S.~Malinov, Z.~Guo, W.~Sha, and A.~Wilson.
\newblock Differential scanning calorimetry study and computer modeling of
  $\beta\rightarrow\alpha$ phase ransformation in a \mbox{Ti-6Al-4V} alloy.
\newblock {\em Metall. Mater. Trans. A}, 32A:879--887, 2001.

\bibitem{Wierszy_1991}
I.~A. Wierszy{\l\l}owski.
\newblock The effect of the thermal path to reach isothermal temperature on
  transformation kinetics.
\newblock {\em Metall. Trans. A}, 22A:993--999, 1991.

\bibitem{Lusk_Jou_1997}
M.~Lusk and H-J Jou.
\newblock On the rule of additivity in phase transformation kinetics.
\newblock {\em Metall. Mater. Trans. A}, 28A:287--291, 1997.

\bibitem{Zhu_et_al_1997}
Y.~T. Zhu, T~C. Lowe, and R.~J. Asaro.
\newblock Assessment of the theoretical basis of the rule of additivity of the
  nucleation incubation time during continuous cooling.
\newblock {\em J. Appl. Phys.}, 82:1129--1137, 1997.

\bibitem{Reti_Felde_1999}
T.~R\'{e}ti and I.~Felde.
\newblock A non-linear extension of additivity rule.
\newblock {\em Comput. Mater. Sci.}, 15:466--482, 1999.

\bibitem{Kampen_et_al_2002}
A.~T.~W. Kempen, F.~Sommer, and E.~J. Mittemeijer.
\newblock Determination and interpretation of isothermal and non-isothermal
  transformation kinetics; the effective activation energies in terms of
  nucleation and growth.
\newblock {\em J. Mater. Sci.}, 37:1321--1332, 2002.

\bibitem{Mittemeijer_Sommer_2002}
E.~J. Mittemeijer and F.~Sommer.
\newblock Solid state phase transformation kinetics: a modular transformation
  model.
\newblock {\em Z. Metallkd.}, 93:352--361, 2002.

\bibitem{Liu_et_al_2004a}
F.~Liu, F.~Sommer, and E.~J. Mittemeijer.
\newblock An analytical model for isothermal and isochronal transformation
  kinetics.
\newblock {\em J. Mater. Sci.}, 39:1621--1634, 2004.

\bibitem{Liu_et_al_2004b}
F.~Liu, F.~Sommer, and E.~J. Mittemeijer.
\newblock Determination of nucleation and growth mechanisms of the
  crystallization of amorphous alloys; application to calorimetric data.
\newblock {\em Acta Mater.}, 52:3207--3216, 2004.

\bibitem{Liu_et_al_2007}
F.~Liu, F.~Sommer, C.~Boss, and E.~J. Mittemeijer.
\newblock Analysis of solid state phase transformation kinetics: models and
  recipes.
\newblock {\em Intern. Mater. Rev.}, 52:193--212, 2007.

\bibitem{Farjas_Roura_2006}
J.~Farjas and P.~Roura.
\newblock Modification of the \mbox{Kolmogorov-Johnson-Mehl-Avrami} rate
  equation for non-isothermal experiments and its analytical solutions.
\newblock {\em Acta Mater.}, 54:5573--5579, 2006.

\bibitem{Elder_et_al_1996}
K.~R. Elder, J.~D. Gunton, and M.~Grant.
\newblock Nonisothermal eutectic crystallization.
\newblock {\em Phys. Rev. B}, 54:6476--6484, 1996.

\bibitem{Cahn_1956a}
J.~W. Cahn.
\newblock The kinetics of grain boundary nucleated reactions.
\newblock {\em Acta Met.}, 4:449--459, 1956.

\bibitem{Massih_et_al_2003}
A.~R. Massih, T.~Andersson, P.~Witt, M.~Dahlb\"{a}ck, and M.~Limb\"{a}ck.
\newblock The effect of quenching rate on the $\beta$-to-$\alpha$ phase
  transformation structure in zirconium alloy.
\newblock {\em J. Nucl. Mater.}, 322:138--151, 2003.

\bibitem{Forgeron_et_al_2000}
T.~Forgeron, J.~C. Brachet, F.~Barcelo, A.~Castaing, J.~Hivroz, J.~P. Mardon,
  and C.~Bernaudat.
\newblock Experiment and modeling of advanced fuel rod cladding under
  \mbox{LOCA} conditions: alpha-beta phase kinetics and \mbox{EDGAR}
  methodology.
\newblock In G.~P. Sabol and G.~D. Moan, editors, {\em Zirconium in Nuclear
  Industry: Twelfth International Symposium}, volume ASTM STP 1345, pages
  256--278, West Conshohocken, PA, USA, 2000. American Society for Testing and
  Materials.

\bibitem{Hermansson_2008}
P.~Hermansson.
\newblock Personal communication, 2008.

\bibitem{Mori_1965a}
H.~Mori.
\newblock Transport, collective motion and \mbox{B}rownian motion.
\newblock {\em Prog. Theor. Phys.}, 33:423--455, 1965.

\bibitem{Forster_1975}
D.~Forster.
\newblock {\em Hydrodynamic Fluctuations, Broken Symmetry, and Correlation
  Functions}.
\newblock W. A. Benjamin, Inc., Reading, MA, USA, 1975.

\bibitem{Mittemeijer_1992}
E.~J. Mittemeijer.
\newblock Review: Analysis of kinetics of phase transformation.
\newblock {\em J. Mater. Sci.}, 27:3977--3987, 1992.

\bibitem{Lemaignan_Motta_1994}
C.~Lemaignan and A.~T. Motta.
\newblock Zirconium alloys in nuclear applications.
\newblock In R.~W. Cahn, P.~Haasen, and E.~J. Kramer, editors, {\em Nuclear
  Materials}, volume 10B of {\em Material Science and Technology}, chapter~7.
  VCH, Weinheim, Germany, 1994.
\newblock Volume editor B.R.T. Frost.

\bibitem{Miquet_et_al_1982a}
A.~Miquet, D.~Charquet, and C.~H. Allibert.
\newblock Solid state phase equilibria of \mbox{Z}ircaloy-4 in the temperature
  range 750-1050$^\circ$\mbox{C}.
\newblock {\em J. Nucl. Mater.}, 105:132--141, 1982.

\bibitem{Bangaru_et_al_1987}
N.~V. Bangaru, R.~A. Busch, and J.~H. Schemel.
\newblock Effects of beta quenching on the microstructure and corrosion of
  \mbox{Z}ircaloys.
\newblock In R.B. Adamson and L.F.P. van Swam, editors, {\em Zirconium in
  Nuclear Industry: Seventh International Symposium}, volume ASTM STP 939,
  pages 341--363, Philadelphia, USA, 1987. American Society for Testing and
  Materials.

\bibitem{Okvist_Kallstrom_1970}
G.~{\"O}kvist and K.~K{\"a}llstr{\"o}m.
\newblock The effect of zirconium carbide on the $\beta\rightarrow\alpha$
  transformation structure in \mbox{Z}ircaloy.
\newblock {\em J. Nucl. Mater.}, 35:316--321, 1970.

\bibitem{Charquet_Alheritiere_1987}
D.~Charquet and E.~Alheritier.
\newblock Influence of the impurities and temperature on the microstructure of
  \mbox{Z}ircaloy-2 and \mbox{Z}ircaloy-4 after $\beta\rightarrow\alpha$
  transformation.
\newblock In R.B. Adamson and L.F.P. van Swam, editors, {\em Zirconium in
  Nuclear Industry: Seventh International Symposium}, volume ASTM STP 939,
  pages 284--291, Philadelphia, USA, 1987. American Society for Testing and
  Materials.

\bibitem{Holt_1970}
R.~A. Holt.
\newblock The beta to alpha phase transformation in \mbox{Z}ircaloy-4.
\newblock {\em J. Nucl. Mater.}, 35:322--334, 1970.

\bibitem{Holt_1973}
R.~A. Holt.
\newblock Comments on the beta to alpha phase transformation in
  \mbox{Z}ircaloy-4.
\newblock {\em J. Nucl. Mater.}, 47:262--264, 1973.

\bibitem{Woo_Tangri_1979}
O.~T. Woo and K.~Tangri.
\newblock Transformation characteristics of rapidly heated and quenched
  \mbox{Z}ircaloy-4-oxygen alloys.
\newblock {\em J. Nucl. Mater.}, 79:82--94, 1979.

\bibitem{Hunt_Schulson_1980}
C.~E.~L. Hunt and E.~M. Schulson.
\newblock Recrystallization of \mbox{Z}ircaloy-4 during transient heating.
\newblock {\em J. Nucl. Mater.}, 92:184--190, 1980.

\bibitem{Corchia_Righini_1981}
M.~Corchia and F.~Righini.
\newblock Kinetic aspects of the phase transformation in \mbox{Z}ircaloy-2.
\newblock {\em J. Nucl. Mater.}, 97:137--148, 1981.

\bibitem{Yoo_Kim_1991}
J.~S. Yoo and I.~S. Kim.
\newblock Effect of ($\alpha+\beta$) heat treatment on the mechanical
  properties of \mbox{Z}ircaloy-4.
\newblock {\em J. Nucl. Mater.}, 185:87--95, 1991.

\bibitem{Brachet_et_al_2002}
J.~C. Brachet, L.~Portier, and T.~Forgeron.
\newblock Influence of hydrogen content on $\alpha/\beta$ phase transformation
  temperatures and on the thermal-mechanical behavior of \mbox{Zy-4},
  \mbox{M4}, and \mbox{M5}$^{TM}$ (\mbox{ZrNbO}) alloys during the first phase
  of \mbox{LOCA} tranient.
\newblock In G.~D. Moan and P.~Rudling, editors, {\em Zirconium in Nuclear
  Industry: Thirteenth International Symposium}, volume ASTM STP 1423, pages
  673--701, West Conshohocken, PA, USA, 2002. American Society for Testing and
  Materials.

\bibitem{Miquet_et_al_1982b}
A.~Miquet, D.~Charquet, C.~Michaut, and C.~H. Allibert.
\newblock Effect of \mbox{Cr}, \mbox{Sn} and \mbox{O} contents on the solid
  state phase boundary temperature of \mbox{Zircaloy-4}.
\newblock {\em J. Nucl. Mater.}, 105:142--148, 1982.

\bibitem{Massih_2009}
A.~R. Massih.
\newblock Transformation kinetics of zirconium alloys under non-isothermal
  conditions.
\newblock {\em J. Nucl. Mater.}, 384:330--335, 2009.

\bibitem{Quarteroni_Saleri_2003}
A.~Quarteroni and F.~Saleri.
\newblock {\em Scientific Computing with MATLAB}.
\newblock Springer, Berlin, Germany, 2003.

\bibitem{Miquet_et_al_1982}
A.~Miquet, D.~Charquet, and C.~H. Allibert.
\newblock Solid-state phase equilibria of \mbox{Z}ircaloy-4 in the temperature
  range 750-1050$^\circ$c.
\newblock {\em J. Nucl. Mater.}, 105:132--141, 1982.

\bibitem{Canay_Danon_Arias_2000}
M.~Canay, C.~A. Danon, and D.~Arias.
\newblock Phase transition temperature in the \mbox{Z}r-rich corner of
  \mbox{Z}r-\mbox{N}b-\mbox{S}n-\mbox{F}e alloys.
\newblock {\em J. Nucl. Mater.}, 280:365--371, 2000.

\bibitem{Zhu_Devletian_1991}
Y.~T. Zhu and J.~H. Devletian.
\newblock Determination of equilibrium solid-phase transition temperature using
  \mbox{DTA}.
\newblock {\em Metall. Trans. A}, 22A:1993--1998, 1991.

\bibitem{Evans_1945}
U.~R. Evans.
\newblock The laws of expanding circles and spheres in relations to the lateral
  growth of surface films and the grain-size of metals.
\newblock {\em Trans. Faraday Soc.}, 41:365--374, 1945.

\bibitem{Jackson_1974}
J.~L. Jackson.
\newblock Dynamics of expanding inhibitory fields.
\newblock {\em Science}, 183:444--446, 1974.

\bibitem{Burbelko_et_al_2005}
A.~A. Burbelko, E.~Fras, and W.~Kapturkiewicz.
\newblock About \mbox{K}olmogorov's statistical theory of phase transformation.
\newblock {\em Mater. Sci. Eng. A}, 413-414:429--434, 2005.

\bibitem{Shneidman_Weinberg_1993}
V.~A. Shneidman and M.~C. Weinberg.
\newblock The effects of transient nucleation on size-dependent growth rate on
  phase transformation kinetics.
\newblock {\em J. Non-Cryst. Solids}, 160:89--98, 1993.

\bibitem{Iwamatsu_2008}
M.~Iwamatsu.
\newblock Direct numerical simulation of homogeneous nucleation and growth in a
  phase-field model using cell dynamics method.
\newblock {\em J. Chem. Phys.}, 128:084504, 2008.

\bibitem{Christian_2002}
J.~W. Christian.
\newblock {\em The Theory of Transformations in Metals and Alloys}.
\newblock Pergamon, Amsterdam, 2002.

\bibitem{Ham_1959}
F.~S. Ham.
\newblock Stress-assisted precipitation on dislocations.
\newblock {\em J. Appl. Phys.}, 30(6):915--926, 1959.

\end{thebibliography}

\end{document}